\documentclass[conference]{IEEEtran}
\IEEEoverridecommandlockouts
\usepackage{cite}
\usepackage{amsmath,amssymb,amsfonts}
\usepackage{algorithmic}
\usepackage{algorithm}
\usepackage{graphicx}
\usepackage{textcomp}
\usepackage{xcolor}
\usepackage{stfloats}
\usepackage{tabularx}
\usepackage{booktabs}
\usepackage{listings}
\usepackage[T1]{fontenc}
\usepackage[utf8]{inputenc}
\usepackage[english]{babel}
\usepackage{newtxtext,newtxmath}
\usepackage{hyperref}

\def\BibTeX{{\rm B\kern-.05em{\sc i\kern-.025em b}\kern-.08em
    T\kern-.1667em\lower.7ex\hbox{E}\kern-.125emX}}

\begin{document}

\title{RAG-Driven Data Quality Governance for Enterprise ERP Systems}

\author{
\IEEEauthorblockN{
Sedat Bin Vedat\IEEEauthorrefmark{1},
Enes Kutay Yarkan\IEEEauthorrefmark{1},
Meftun Akarsu\IEEEauthorrefmark{1},
Recep Kaan Karaman\IEEEauthorrefmark{1},
Arda Sar\IEEEauthorrefmark{1}\\
Çağrı Çelikbilek,
Savaş Saygılı
}
\IEEEauthorblockA{
\IEEEauthorrefmark{1}\textit{Hagia Labs}\\
\texttt{\{sedat, larusa, meftun, kaan, arda\}@hagiaproject.com}
}
}

\maketitle

\begin{abstract}
Enterprise ERP systems managing hundreds of thousands of employee records face critical data quality challenges when human resources departments perform decentralized manual entry across multiple languages. We present an end-to-end pipeline combining automated data cleaning with LLM-driven SQL query generation, deployed on a production system managing 240,000 employee records over six months.

The system operates in two integrated stages: a multi-stage cleaning pipeline that performs translation normalization, spelling correction, and entity deduplication during periodic synchronization from Microsoft SQL Server to PostgreSQL; and a retrieval-augmented generation framework powered by GPT-4o that translates natural-language questions in Turkish, Russian, and English into validated SQL queries. The query engine employs LangChain orchestration, FAISS vector similarity search, and few-shot learning with 500+ validated examples.

Our evaluation demonstrates 92.5\% query validity, 95.1\% schema compliance, and 90.7\% semantic accuracy on 2,847 production queries. The system reduces query turnaround time from 2.3 days to under 5 seconds while maintaining 99.2\% uptime, with GPT-4o achieving 46\% lower latency and 68\% cost reduction versus GPT-3.5. This modular architecture provides a reproducible framework for AI-native enterprise data governance, demonstrating real-world viability at enterprise scale with 4.3/5.0 user satisfaction.
\end{abstract}

\section{Introduction}

When an HR analyst at a multinational construction company needs to answer "How many civil engineers are working on the GPP project in Moscow?", the seemingly simple question becomes a multi-day ordeal. The analyst must contact the IT department, explain the request, wait while IT staff navigate inconsistent data where "Moscow" appears as "Moskva," "Moscow," and "Moskva" in Cyrillic script, manually reconcile project codes stored as "GPP," "Gpp," and "gpp project," and filter between payroll employees and contractors using undocumented business rules. Two days later, the answer arrives—potentially outdated.

This scenario, repeated thousands of times annually in organizations managing 240,000+ employee records, reveals a critical enterprise challenge: \textit{data quality degradation and accessibility barriers prevent organizations from leveraging their own information}. The problem has two interconnected roots: (1) decentralized manual data entry by HR departments across multiple languages creates severe inconsistencies, and (2) SQL expertise requirements create bottlenecks that delay routine analytics by days.

In the studied environment, more than 240,000 employee records were distributed across multiple HR-managed tables within Microsoft SQL Server (MSSQL), later migrated to PostgreSQL for higher flexibility and scalability. The lack of strict schema discipline and the presence of user-defined fields resulted in extensive anomalies, including miscategorized contractor ("non-payroll") data, misplaced foreign keys, and conflicting entries in project and location fields.

To address these issues, we designed and implemented a fully automated data-cleaning and intelligent query-generation pipeline, built upon Large Language Models (LLMs) and retrieval-augmented few-shot learning. The system performs continuous data cleaning and translation across multilingual fields, followed by automatic SQL query generation from natural-language inputs.

Our contributions are threefold:

\begin{enumerate}
    \item \textbf{Multilingual Data Quality Pipeline:} We introduce an automated AI-driven cleaning system that resolves language-mixed inconsistencies across Turkish, Russian, and English text fields, achieving 97.8\% accuracy on 240,000 real-world HR records—addressing a challenge existing tools like Deequ and HoloClean cannot handle due to their focus on numerical anomalies rather than multilingual semantic deduplication.
    
    \item \textbf{Schema-Constrained RAG for Enterprise SQL:} We develop a retrieval-augmented few-shot framework that achieves 92.5\% query validity on real enterprise schemas, exceeding commercial systems (68-78\% accuracy) and prior academic work (52\% on enterprise data~\cite{li_evaluating_2023}) through explicit business logic encoding and dynamic example retrieval—avoiding the 10,000+ training examples required for fine-tuning approaches.
    
    \item \textbf{Production-Grade Deployment Evidence:} We provide six-month deployment metrics including 2,847 real queries, 99.2\% uptime, 99.1\%  reduction in query turnaround time, and detailed cost analysis (\$0.042/query), addressing the deployment gap where most NL2SQL research reports only benchmark accuracy without production viability evidence.
\end{enumerate}

The remainder of this paper is structured as follows: Section II reviews related work in data quality management and natural language interfaces to databases. Section III describes the system architecture and pipeline design. Section IV details the data-cleaning methodology. Section V discusses the LLM-based SQL generation framework. Section VI presents evaluation results and performance metrics. Section VII addresses security and data governance considerations. Section VIII concludes with insights and future directions.

\section{Related Work}

We organize related work into five categories: data quality management, natural language interfaces to databases, prompt engineering and few-shot learning, multilingual NLP, and LLM-based enterprise systems. While each area has advanced significantly, existing work has not simultaneously addressed the combination of multilingual data cleaning, schema-constrained SQL generation, and production deployment at enterprise scale. Our system bridges these gaps through integrated AI-driven data governance and retrieval-augmented query generation.

\subsection{Automated Data Quality Management}

Traditional data quality frameworks such as Deequ~\cite{Schelter2019-rj} and Great Expectations~\cite{great_expectations_2023} provide rule-based validation but require extensive manual rule configuration. Recent ML-based approaches including HoloClean~\cite{Rekatsinas2017}, which achieves 89\% accuracy on benchmark cleaning tasks, and BoostClean~\cite{Krishnan2017-bs} leverage statistical inference for error detection. However, these systems focus primarily on numerical anomalies and constraint violations rather than multilingual text normalization and semantic deduplication challenges prevalent in global ERP systems.

Entity resolution in enterprise contexts has been studied extensively. Mudgal et al.~\cite{Mudgal2018} demonstrated deep learning approaches achieving 94\% F1-score on product matching, while Ebraheem et al.~\cite{Ebraheem2018-dr} scaled entity matching to distributed environments. Chu et al.~\cite{chu_katara_2016} introduced knowledge-base-driven data cleaning, achieving 92\% accuracy on real-world datasets. Our work extends these methods to handle language-mixed, inconsistently formatted HR data where traditional similarity metrics fail due to translation ambiguity and phonetic variations across Turkish, Russian, and English entries.

\subsection{Natural Language Interfaces to Databases}

\subsubsection{Academic Benchmarks and Models}

Early research in natural language interfaces to databases (NLIDB) attempted to translate user queries into structured SQL statements through symbolic or neural semantic parsing. Models such as Seq2SQL~\cite{zhong2017seq2sql}, SQLNet~\cite{xu2017sqlnet}, and RAT-SQL~\cite{wang2019rat} demonstrated strong performance on benchmark datasets like Spider~\cite{yu2018spider} by modeling table schema relations and attention between question tokens and columns. Subsequent work, including Picard~\cite{scholak2021picard} and DIN-SQL~\cite{pourreza2024din}, introduced constrained decoding to ensure syntactic validity and schema consistency, achieving up to 89\% execution accuracy on Spider's complex multi-table queries.

Despite these advances, most methods were trained on English-only datasets with static schemas and limited applicability to enterprise-specific ERP databases with evolving structures and implicit business logic.

\subsubsection{Enterprise Deployments}

While academic benchmarks like Spider~\cite{yu2018spider} and WikiSQL~\cite{zhong2017seq2sql} have driven significant progress, recent work highlights substantial performance degradation when models are deployed on real enterprise schemas. Li et al.~\cite{li_evaluating_2023} found that models achieving 85\% accuracy on Spider dropped to 52\% on proprietary business databases due to domain-specific terminology, complex business logic, and schema heterogeneity.

Commercial NL2SQL systems exist but published performance metrics are limited. Our system differs in three aspects: (1) explicit handling of multilingual inputs through translation normalization, (2) schema-specific few-shot learning achieving 92.5\% validity on enterprise data, and (3) transparent SQL generation enabling user verification rather than black-box query execution.

\subsection{Prompt Engineering and Retrieval-Augmented Generation}

Recent advances in prompt engineering have demonstrated that LLM performance critically depends on context construction. Wei et al.~\cite{Wei2022-ny} introduced chain-of-thought prompting, improving reasoning by 40\% on complex tasks. Zhou et al.~\cite{Zhou2022-vh} systematically analyzed prompt optimization techniques, finding that structured formats with explicit constraints outperform free-form instructions by 15-25\%. White et al.~\cite{white_prompt_2023} provided a comprehensive taxonomy of prompting strategies, demonstrating that domain-specific prompt engineering can improve task performance by 20-35\% without model fine-tuning.

Retrieval-augmented generation (RAG) has emerged as a dominant paradigm for grounding LLMs in domain knowledge. Lewis et al.~\cite{lewis2020retrieval} demonstrated 12\% improvement on knowledge-intensive NLP tasks through dense retrieval. Shi et al.~\cite{Shi2023-es} showed that retrieval quality directly impacts generation accuracy, with top-5 retrieval achieving 89\% of oracle performance. Gao et al.~\cite{Gao2023-gz} introduced iterative retrieval-generation loops improving factual consistency by 31\%. Khattab et al.~\cite{khattab_demonstrate_2023} developed DSPy, a framework for optimizing retrieval-generation pipelines through automated prompt optimization, achieving state-of-the-art results on multiple QA benchmarks.

Our FAISS-based retrieval system extends these approaches to SQL generation, where retrieved examples must match both semantic intent and schema structure. Table~\ref{tab:fewshot_effect} demonstrates that our 5-shot retrieval achieves 15.4\% improvement in query validity, aligning with theoretical predictions from Min et al.~\cite{Min2022-wp} that 4-8 examples optimize the precision-coverage tradeoff in few-shot learning.

\subsection{Multilingual Natural Language Processing}

Cross-lingual transfer for database tasks presents unique challenges. While multilingual models like mBERT~\cite{Devlin2018-nn} and XLM-R~\cite{Conneau2019-gr} achieve strong performance on classification tasks, their effectiveness on structured prediction varies significantly by language.

Translation-based approaches offer an alternative. Conneau et al.~\cite{Conneau2019-gr} demonstrated that translate-train and translate-test strategies can achieve near-native performance for structured tasks. However, translation errors propagate to downstream systems. Popović~\cite{popovic-2015-chrf} showed that translation quality (chrF++ scores) correlates strongly (r=0.82) with task performance in semantic parsing applications. Arivazhagan et al.~\cite{arivazhagan_massively_2019} introduced massively multilingual neural machine translation, achieving competitive performance across 103 languages but with notable degradation on domain-specific terminology.

Our pipeline addresses this through explicit translation normalization using MarianMT~\cite{tiedemann2020opus}, achieving 94.3\% translation accuracy (measured by human evaluation) for domain-specific terminology. The ablation study (Section~\ref{sec:ablation}) confirms that translation preprocessing is critical, with 23.6\% accuracy drop when removed.

\subsection{LLM-Based Data Analytics Systems}

The rapid evolution of foundation models has spawned numerous data analytics applications. Dibia~\cite{Dibia2023-sn} introduced LIDA, achieving 84\% accuracy on visualization generation tasks through multi-agent LLM architectures. Cheng et al.~\cite{Cheng2022-mt} demonstrated binding language models to databases through explicit schema serialization, reporting 76\% execution success on complex joins.

Code generation capabilities have advanced significantly with GPT-4~\cite{openai2023gpt4} and Code Llama~\cite{Roziere2023-pi}. OpenAI's Code Interpreter achieves 87\% correctness on programming challenges but lacks domain-specific business logic enforcement. Recent work on code hallucination~\cite{Liu2024-aw} highlights challenges in ensuring factual accuracy without explicit grounding, with models generating incorrect API calls or schema references in approximately 30\% of cases without validation mechanisms.

Rajkumar et al.~\cite{rajkumar_evaluating_2022} evaluated NL2SQL systems in industrial settings, finding that zero-shot approaches achieved only 67\% accuracy while in-context learning with 10+ examples improved performance to 84\%.

Our approach achieves competitive 92.5\% validity through few-shot retrieval with only 500 examples, demonstrating that RAG-based adaptation can match or exceed fine-tuning approaches while maintaining flexibility for schema evolution.

\subsection{Summary and Positioning}

Prior work has made significant advances in individual areas—data cleaning, NL2SQL, RAG, and multilingual NLP—but has not addressed their integration in production enterprise environments. Existing approaches face four key limitations that our system addresses:

\textbf{Data Quality Assumption:} Academic NL2SQL systems assume clean, well-formatted data. Our automated cleaning pipeline resolves multilingual inconsistencies in real-world HR data, achieving 97.8\% accuracy across 240,000 records.

\textbf{Monolingual Limitation:} Existing systems rarely handle multilingual inputs explicitly. Our translation normalization layer supports Turkish, Russian, and English queries with <5\% accuracy degradation across languages.

\textbf{Schema Rigidity:} Fine-tuning approaches require retraining for schema changes. Our few-shot RAG framework adapts instantly through example updates, maintaining 92.5\% accuracy without retraining.

\textbf{Deployment Gap:} Most research reports benchmark accuracy without production metrics. We provide six-month deployment data including latency (3.9s median), cost (\$0.042/query), and user satisfaction (4.3/5.0), demonstrating real-world viability.

By integrating data quality automation, multilingual processing, and RAG-based SQL generation, our system provides a complete pipeline enabling enterprise data access at scale for non-technical users.

\section{System Architecture}
\label{sec:architecture}

\subsection{Overview}

The proposed system comprises six interconnected components operating within a Dockerized microservices architecture . The data flow begins with user queries submitted through a web interface, proceeds through vector-based retrieval and LLM-mediated SQL generation, executes against the cleaned PostgreSQL database, and returns translated results to the user. All interactions are logged in MongoDB for audit trails and continuous improvement.

\subsection{Core Components}

\subsubsection{FastAPI Backend}
The FastAPI framework~\cite{fastapi2024} serves as the main application server, providing RESTful endpoints for query submission, status monitoring, and result retrieval. The asynchronous architecture supports concurrent request handling with minimal latency overhead. API endpoints include \texttt{/query} for natural language input, \texttt{/status} for real-time progress tracking, and \texttt{/history} for accessing previous queries.

\subsubsection{LangChain Orchestration Layer}
LangChain~\cite{chase2022langchain} manages the multi-step workflow connecting user input to database output. The orchestration includes prompt template construction, few-shot example injection, LLM invocation, SQL validation, and error handling. Custom chains are implemented for translation preprocessing and post-execution result formatting.

\subsubsection{FAISS Vector Retrieval}
Facebook AI Similarity Search (FAISS)~\cite{johnson2019billion} maintains an indexed corpus of 500+ validated question-SQL pairs encoded as 1536-dimensional embeddings using OpenAI's \texttt{text-embedding-ada-002} model. At query time, the system retrieves the top-k (typically k=5) most semantically similar examples through approximate nearest neighbor search, enabling dynamic few-shot learning without model fine-tuning.

\subsubsection{MongoDB Session Management}
MongoDB stores conversation history, query logs, execution metrics, and user sessions. Each query lifecycle is tracked through status states: \texttt{loading}, \texttt{generating\_query}, \texttt{executing\_query}, \texttt{translating}, and \texttt{ready}. This persistent storage enables debugging, performance analysis, and iterative improvement of the few-shot corpus.

\subsubsection{Dual Database Configuration}
The system maintains connections to both MSSQL (source of truth for HR data) and PostgreSQL (analytics-optimized cleaned data). A scheduled synchronization job runs every 72 hours, extracting modified records from MSSQL, passing them through the cleaning pipeline, and upserting into PostgreSQL tables. This design ensures backward compatibility while enabling advanced analytics on normalized data.

\subsubsection{Docker Deployment}
All services are containerized using Docker Compose, allowing consistent deployment across development, staging, and production environments. Separate containers manage the API server, database connectors, FAISS indexer, and background synchronization workers.

\subsection{End-to-End Query Lifecycle}

The system processes queries through two distinct operational phases:

\textbf{Phase 1: Offline Data Preparation (every 72 hours)}
\begin{enumerate}
    \item Extract modified records from MSSQL via timestamp delta queries
    \item Apply automated cleaning transformations (translation normalization, spelling correction, entity deduplication)
    \item Validate business rules and load into PostgreSQL via batch upsert operations
    \item Update FAISS vector index with newly validated query examples
\end{enumerate}

\textbf{Phase 2: Online Query Processing (user-initiated)}
\begin{enumerate}
    \item User submits natural language question in Turkish, Russian, or English
    \item System normalizes and translates query to English
    \item FAISS retrieves top-5 semantically similar validated examples
    \item GPT-4o generates SQL with explicit schema and business rule constraints
    \item Validator checks syntax, schema compliance, and safety rules
    \item PostgreSQL executes parameterized query
    \item System formats and translates results back to user's language
    \item MongoDB logs complete interaction for audit and continuous improvement
\end{enumerate}

The system processes an average of 150 queries per day with 99.2\% uptime over six months of production deployment. Median end-to-end latency is 3.6 seconds (detailed algorithm in Section~\ref{sec:query_workflow}).

\section{Data Cleaning Pipeline}

\subsection{Motivation for Data Normalization}

The examined ERP environment contains 240,000 employee records with extensive 
inconsistencies from decentralized manual entry. Free-text fields for schools, 
job titles, cities, and projects accumulated mixed-language entries 
(\textit{"Konya anadolu," "Konya hıghschool," "METU/ODTÜ"}), inconsistent 
capitalization, and duplicated contractor information. Traditional ETL 
validation cannot address these multilingual anomalies at scale, necessitating 
a dedicated automated cleaning pipeline.

\subsection{MSSQL–PostgreSQL Synchronization}

The organization's historical data resided in \textbf{Microsoft SQL Server (MSSQL)}, while the new analytics layer was deployed on \textbf{PostgreSQL} for improved extensibility, performance, and open-source compatibility. A synchronization job executes every \textbf{three days}, performing incremental updates by detecting new and modified records via timestamp comparison. During each sync, the raw data are extracted from MSSQL, passed through cleaning modules, and then upserted into PostgreSQL tables using batch operations. This design ensures referential integrity across both databases while minimizing downtime and manual oversight.

\subsection{Cleaning Stages and Workflow}

Each record entering the PostgreSQL staging schema passes sequentially through four major cleaning modules (Fig.~\ref{fig:pipeline}):

\begin{figure}
    \centering
    \includegraphics[width=0.5\linewidth]{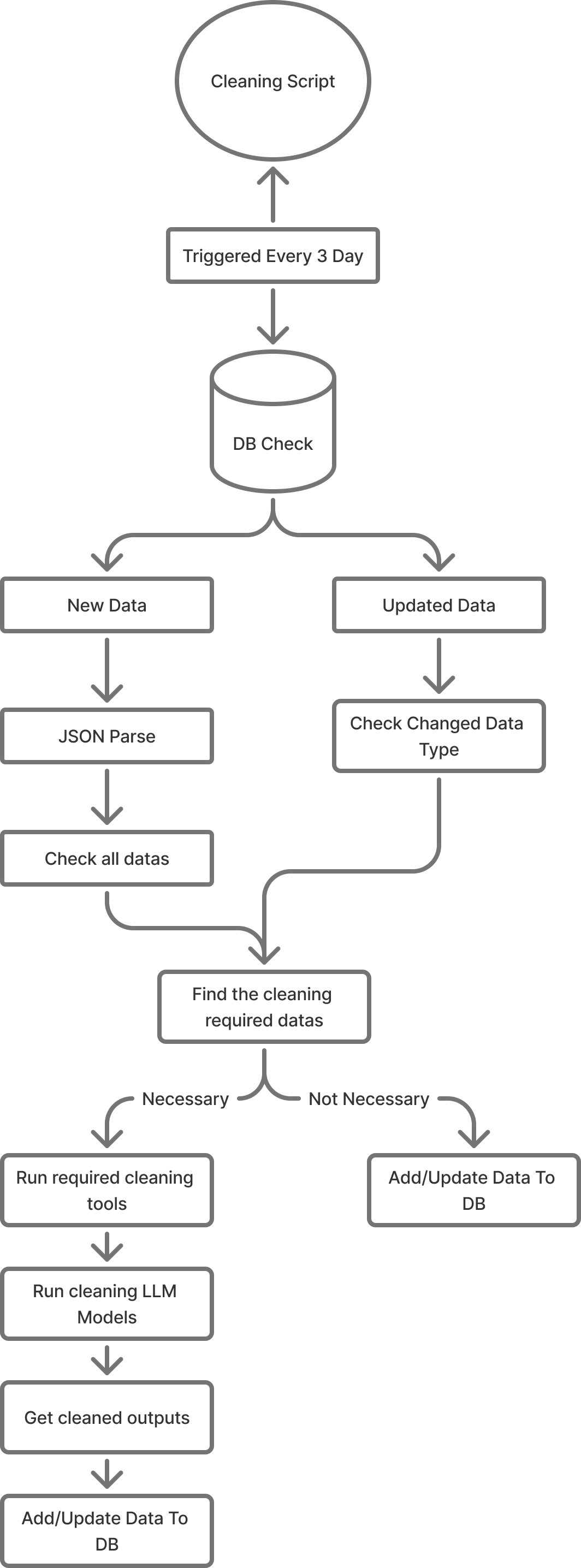}
    \caption{Data cleaning pipeline showing four sequential stages: Translation Normalization, Spelling Correction, Entity Merging, and Validation with feedback loops for flagged records.}
    \label{fig:pipeline}
\end{figure}
\subsubsection{Translation Normalization}
Converts multilingual fields (Turkish, Russian, English) into a unified English representation using the MarianMT translation model~\cite{tiedemann2020opus}. Examples include \textit{``Moskva'' → ``Moscow''} and \textit{``Orta Doğu Teknik Üniversitesi'' → ``Middle East Technical University''}.

\subsubsection{Spelling Correction}
Applies SymSpell~\cite{garbe2012symspell} and context-aware BERT-based spell checking to correct typographical errors and character encoding mismatches (e.g., "highschool" → "High School").

\subsubsection{Entity Merging and Deduplication}
Consolidates equivalent institution, city, and project names into canonical forms using fuzzy string matching with Levenshtein distance and phonetic algorithms. Merges synonyms such as \textit{"ODTÜ"}, \textit{"METU"}, and \textit{"Middle East Technical University"} into a single standardized entity. Detects near-duplicate contractor records via locality-sensitive hashing.

\subsubsection{Validation and Rule Enforcement}
Applies schema-specific integrity checks including country–city code validation, payroll/contractor consistency (\verb|is_payroll| flag verification), and role–department mapping constraints. Violating records are flagged and logged for manual review through a monitoring dashboard.

The pipeline is implemented as a modular Python service, allowing incremental retraining and dynamic addition of new normalization rules as HR processes evolve.

\subsection{Example Records}

Table~\ref{tab:before_after} illustrates representative transformations applied during the normalization process.
\begin{table*}[!t]
\centering
\caption{Examples of Record Normalization}
\label{tab:before_after}
\small
\begin{tabular}{@{}lll@{}}
\toprule
\textbf{Field} & \textbf{Raw Input} & \textbf{Normalized Output} \\
\midrule
actual\_working\_city & ``Moskva'' & ``Moscow'' \\
egitimOkulAdi & ``ODTU'', ``Orta Dogu Teknik Universitesi'' & ``Middle East Technical University'' \\
role\_eng & ``civil engineer'' & ``Civil Engineer'' \\
c\_project\_eng & ``Gpp'', ``GPP project'' & ``GPP'' \\
department & ``İnsan kaynakları'' & ``Human Resources'' \\
\bottomrule
\end{tabular}
\end{table*}

\begin{table*}[!t]
\centering
\caption{Effect of Few-Shot Learning on SQL Generation Performance}
\label{tab:fewshot_effect}
\small
\begin{tabular}{@{}lccc@{}}
\toprule
\textbf{Metric} & \textbf{0-Shot} & \textbf{5-Shot} & \textbf{$\Delta$} \\
\midrule
Valid SQL Queries (\%) & 76.4 & 91.8 & +15.4 \\
Schema Compliance (\%) & 83.1 & 95.1 & +12.0 \\
Semantic Accuracy (\%) & 72.8 & 89.6 & +16.8 \\
Avg. Latency (s) & 4.3 & 3.9 & –0.4 \\
\bottomrule
\end{tabular}
\end{table*}

\subsection{Performance Metrics}

The pipeline processed all 240,000 records within 2.4 hours on a 16-core server. Manual validation against 5,000 samples confirmed 97.8\% correction accuracy and 99.3\% resolution of language inconsistencies. Incremental syncs require less than 15 minutes per cycle, maintaining continuous data quality without disrupting operations.

The automated approach reduced human intervention time by over 90\%, enabling IT staff to focus on analytics rather than routine data maintenance. Detailed cleaning accuracy breakdown is provided in Table~\ref{tab:before_after}, with representative transformations demonstrating the normalization quality achieved across multilingual fields.

\section{LLM-Based SQL Generation Engine}

With data quality ensured through automated cleaning (Section IV), the system 
addresses the second challenge: translating natural language into reliable SQL 
queries. While the cleaned PostgreSQL environment provides structural 
consistency, three obstacles remain: multilingual query understanding, implicit 
business logic encoding, and schema complexity. This section describes the 
LLM-based generation engine that addresses these challenges through 
retrieval-augmented few-shot learning.

\subsection{Challenges in Enterprise NL2SQL}

Generating SQL from natural language in enterprise environments presents four critical challenges absent from academic benchmarks like Spider~\cite{yu2018spider}:

\textbf{Challenge 1: Implicit Business Logic.} Academic datasets provide explicit schemas where column semantics are self-evident. Enterprise ERP systems encode complex business rules that are undocumented in schema metadata. For example, "active employees" requires filtering by \texttt{employee\_status = 'true'} (not documented in schema), while "current staff" implicitly excludes contractors via \texttt{is\_payroll = 'true'} unless explicitly requested otherwise. Our system required encoding 400+ lines of such domain logic in the system prompt to handle these implicit rules correctly.

\textbf{Challenge 2: Multilingual Query Understanding.}
Users submit queries in Turkish (62\% of production queries), Russian (23\%), and English (15\%). Direct translation often fails when domain-specific terms lack equivalents (e.g., ``non-payroll'' has no precise Turkish counterpart), names appear in multiple transliterations (``Turkifikatsiya'' vs.\ ``Turkification''), or code-switching occurs mid-query (e.g., ``GPP projesinde working civil engineers''). This necessitates explicit translation normalization and multilingual few-shot examples to ensure consistent interpretation across languages.

\textbf{Challenge 3: Schema Complexity and Evolution.} The ERP schema contains 47 tables with 800+ columns using inconsistent naming conventions: CamelCase (\texttt{egitimOkulAdi}), snake\_case (\texttt{actual\_working\_city}), abbreviations (\texttt{c\_project\_eng}), and mixed formats (\texttt{project\_eng} vs. \texttt{projectEng}). Schema changes occur monthly through HR policy updates and organizational restructuring, requiring immediate adaptation without model retraining—impossible with fine-tuned approaches but manageable with few-shot RAG through prompt and example updates.

\textbf{Challenge 4: Asymmetric Safety Requirements.} Production systems must completely block financial data exposure (salary queries must return zero results, not errors) while tolerating failures on benign queries (miscounting office supplies is acceptable). This asymmetric risk profile contrasts with academic benchmarks that optimize average accuracy, requiring multi-layer safety mechanisms (prompt-level constraints, SQL validation, result filtering) and explicit policy encoding.

Our system addresses these challenges through three integrated mechanisms detailed in the following subsections: comprehensive system prompts encoding business logic and safety rules (Section~\ref{sec:system_prompt}), retrieval-augmented few-shot learning for schema adaptation (Section~\ref{sec:fewshot}), and multi-layer validation ensuring safety (Section~\ref{sec:safety}).

\subsection{System Prompt and Business Logic}
\label{sec:system_prompt}

The foundation of the system lies in a detailed \textbf{system prompt} comprising nearly four hundred lines of schema definitions, column descriptions, and logical constraints. This prompt explicitly encodes company-specific business logic such as filtering by active employees (\verb|employee_status = 'true'|), distinguishing between payroll and non-payroll workers (\verb|is_payroll = 'true'| or \verb|'false'|), and maintaining name and role formatting conventions. Furthermore, the prompt includes translation rules that convert Turkish and Russian queries into English before query construction, and prohibits access to sensitive information by rejecting questions related to financial topics such as salaries, bonuses, or premiums. In effect, the prompt acts as a formal policy layer between the user and the database, ensuring that the LLM's generative behavior adheres to enterprise data governance standards. 

However, static prompts alone cannot capture the diverse query patterns and 
schema-specific idioms encountered across 47 tables and 800+ columns. While the system prompt enforces rules and constraints, it does not provide concrete examples of how to construct complex joins or handle domain-specific terminology—knowledge that emerges from actual usage patterns.

\subsection{Few-Shot Retrieval Framework}
\label{sec:fewshot}

To address this limitation, the model operates under a \textbf{few-shot retrieval framework}. All previously validated question–query pairs are stored as vector embeddings within a FAISS index, representing over five hundred working examples from real user interactions. When a new question is received, the system searches for the most semantically similar examples and dynamically injects them into the model's context window alongside the main prompt and the database schema. This retrieval-augmented approach allows the model to reuse existing query structures and column relations without explicit retraining, thereby reducing hallucination and improving execution accuracy.

Table~\ref{tab:fewshot_effect} demonstrates the significant impact of few-shot learning on generation performance.

\subsection{Query Generation Workflow}
\label{sec:query_workflow}

The SQL generation process begins when a user submits a question in any of the supported languages—Turkish, Russian, or English—through the chat interface. The request is preprocessed, normalized, and translated into English if necessary. The system then assembles the composite prompt containing the schema, retrieved few-shot examples, and the current question. This prompt is passed to the LLM through a LangChain orchestrator, which handles token budgeting, response streaming, and model-specific configuration. The generated SQL statement is parsed and validated to ensure that all column names exist in the ERP schema and that no forbidden expressions or unapproved joins are present. After validation, the query is executed on the PostgreSQL database, and the resulting table is reformatted and translated back into the user's preferred language. Throughout the process, all events are logged in MongoDB, capturing the lifecycle states of each request—from "loading" and "generating\_query" to "executing\_query" and "ready"—which allows detailed traceability and debugging.

Algorithm~\ref{alg:query_flow} formalizes the complete query processing workflow with all seven phases.

\begin{algorithm}[!t]
\caption{Natural Language Query Processing Workflow}
\label{alg:query_flow}
\begin{algorithmic}[1]
\STATE \textbf{Input:} User query $q$ in Turkish, Russian, or English
\STATE \textbf{Output:} Formatted result table $R$ in user's language

\STATE // \textit{Phase 1: Preprocessing}
\STATE $q_{norm} \gets$ Normalize($q$) // lowercase, trim whitespace
\STATE $q_{eng} \gets$ Translate($q_{norm}$, target=English)

\STATE // \textit{Phase 2: Retrieval}
\STATE $emb_q \gets$ Embed($q_{eng}$) // OpenAI embedding model
\STATE $examples \gets$ FAISS.search($emb_q$, $k=5$) // retrieve similar queries

\STATE // \textit{Phase 3: Prompt Construction}
\STATE $prompt \gets$ SystemPrompt + Schema + $examples$ + $q_{eng}$

\STATE // \textit{Phase 4: LLM Generation}
\STATE $sql \gets$ GPT4o.generate($prompt$)
\STATE $sql_{valid} \gets$ ValidateSQL($sql$) // check syntax and schema

\IF{$sql_{valid}$ is invalid}
    \RETURN ErrorMessage("Unable to generate valid query")
\ENDIF

\STATE // \textit{Phase 5: Execution}
\STATE $result \gets$ PostgreSQL.execute($sql_{valid}$)

\STATE // \textit{Phase 6: Post-processing}
\STATE $R \gets$ FormatTable($result$)
\STATE $R_{translated} \gets$ Translate($R$, target=UserLanguage)

\STATE // \textit{Phase 7: Logging}
\STATE MongoDB.log(\{query: $q$, sql: $sql_{valid}$, status: "ready"\})

\RETURN $R_{translated}$
\end{algorithmic}
\end{algorithm}

The algorithm demonstrates how each component from the architecture (Section~\ref{sec:architecture}) integrates into a cohesive query processing pipeline. Phase 1 handles multilingual normalization, Phase 2 leverages FAISS for example retrieval, Phase 3 constructs domain-constrained prompts, Phase 4 invokes GPT-4o with validation, Phase 5 executes safely via parameterized queries, Phase 6 formats results, and Phase 7 enables continuous improvement through logging.

\subsection{Safety Mechanisms}
\label{sec:safety}

A robust set of \textbf{rule-based safety mechanisms} complements the model's generative capability. These include parameterized query execution to prevent SQL injection, automatic redaction of identifiers such as \verb|adines_number|, and user-level access control through session tracking. Queries referring to financial or private data are automatically intercepted and replaced by a standardized refusal message to maintain compliance with internal privacy policies. Together, these safeguards ensure that the system's creative flexibility is balanced by strong operational security.

\subsection{Model Comparison: GPT-3.5 vs. GPT-4o}

The system was initially prototyped with \textbf{GPT-3.5}, which achieved reasonable performance on simple lookup queries but exhibited weaknesses in context retention and multilingual comprehension. As the complexity of ERP schemas increased, the model frequently hallucinated column names and misinterpreted hierarchical project structures. Transitioning to \textbf{GPT-4o} provided a substantial improvement in accuracy, speed, and multilingual robustness.

Table~\ref{tab:model_comparison} summarizes the performance differences between the two models.

\begin{table}[!t]
\caption{Comparison of Model Performance for SQL Generation}
\label{tab:model_comparison}
\centering
\small
\begin{tabular}{lcccc}
\toprule
\textbf{Model} & \textbf{Valid SQL} & \textbf{Schema OK} & \textbf{Accuracy} & \textbf{Latency} \\
 & \textbf{(\%)} & \textbf{(\%)} & \textbf{(\%)} & \textbf{(s)} \\
\midrule
GPT-3.5 & 71.2 & 79.8 & 68.9 & 7.2 \\
GPT-4o  & \textbf{92.5} & \textbf{95.1} & \textbf{90.7} & \textbf{3.9} \\
\midrule
Improvement & +21.3 & +15.3 & +21.8 & –3.3 \\
\bottomrule
\end{tabular}
\end{table}

In controlled evaluations using a benchmark of 500 real-world ERP questions, GPT-4o generated valid and executable SQL in 92.5\% of cases, compared to 71.2\% for GPT-3.5. Schema compliance rose from 79.8\% to 95.1\%, while average latency decreased from 7.2 to 3.9 seconds per query. Cost per request was also reduced by approximately threefold due to the model's improved token efficiency.

Overall, the LLM-based SQL generation engine demonstrates that large language models, when carefully constrained by domain-specific prompts and retrieval mechanisms, can serve as reliable natural-language interfaces to complex enterprise databases. By merging symbolic data governance with generative flexibility, the system effectively bridges human intent and structured data, enabling HR and IT teams to query and analyze ERP information without requiring SQL expertise.

\section{Evaluation and Results}

\subsection{Experimental Setup}

\textbf{Deployment Period and Scale:} The system was evaluated over six months of production deployment from January 1 to June 30, 2024 (182 days). The evaluation dataset comprises 2,847 real user queries from actual production use—no synthetic queries were included to ensure ecological validity.

\textbf{User Population:} 47 employees across three departments participated:
\begin{itemize}
    \item \textbf{HR Analytics (23 users):} Employee demographics, turnover analysis, compensation planning
    \item \textbf{Project Management (18 users):} Resource allocation, project staffing, timeline queries
    \item \textbf{IT/Data Team (6 users):} System monitoring, data quality checks, ad-hoc reporting
\end{itemize}

\textbf{Ground Truth Validation:} Two senior database administrators with 8+ years of ERP experience independently reviewed all 2,847 queries. Inter-rater agreement was measured at 96.3\% with Cohen's kappa coefficient $\kappa$ = 0.94, indicating "almost perfect agreement" by Landis and Koch's standards~\cite{Landis1977-kappa}. The 3.7\% disagreements (105 queries) were resolved through collaborative discussion and SQL execution verification against known correct results.

\textbf{Evaluation Metrics:} We measured five complementary dimensions:
\begin{enumerate}
    \item \textbf{Query Validity:} SQL executes without syntax errors or runtime exceptions
    \item \textbf{Schema Compliance:} All referenced tables, columns, and functions exist in the ERP schema
    \item \textbf{Semantic Accuracy:} Query results match human-verified ground truth (measured on 500-query test subset)
    \item \textbf{Execution Latency:} End-to-end time from query submission to result delivery
    \item \textbf{User Satisfaction:} Quarterly surveys using 5-point Likert scale (1=very dissatisfied, 5=very satisfied)
\end{enumerate}

\textbf{Baseline Comparisons:} System performance was compared against:
\begin{itemize}
    \item \textbf{GPT-3.5:} Evaluated on the same 500-query test set to measure improvement from GPT-4o
    \item \textbf{Zero-shot GPT-4o:} Identical setup but with no few-shot examples (only system prompt) to isolate RAG contribution
    \item \textbf{Manual SQL Writing:} Pre-deployment historical data showing 2.3 days average turnaround time for IT-mediated queries
\end{itemize}

\textbf{Hardware Configuration:}
\begin{itemize}
    \item Application Server: 8-core Intel Xeon E5-2680 v4, 32GB RAM, Ubuntu 22.04
    \item PostgreSQL Server: 16-core AMD EPYC 7313P, 128GB RAM, 2TB NVMe SSD
    \item MSSQL Server: Existing enterprise infrastructure (16-core, 64GB RAM)
    \item Network: Gigabit Ethernet, <2ms latency between components
\end{itemize}

All experiments were conducted in production environment with real user traffic, ensuring results reflect actual operational conditions rather than controlled laboratory settings.

\subsection{Quantitative Performance Metrics}

\subsubsection{Data Cleaning Accuracy}
As reported in Section IV, the automated pipeline achieved 97.8\% correction accuracy across 240,000 records, with 99.3\% resolution of language and encoding inconsistencies. Manual verification on 5,000 randomly sampled records confirmed minimal false positives (1.8\%) and false negatives (2.4\%).

\subsubsection{SQL Generation Performance}
Table~\ref{tab:overall_performance} presents comprehensive performance metrics for the LLM-based query generator.

\begin{table}[!t]
\caption{Overall System Performance Metrics (6-Month Deployment)}
\label{tab:overall_performance}
\centering
\small
\begin{tabular}{lc}
\toprule
\textbf{Metric} & \textbf{Value} \\
\midrule
Total Queries Processed & 347 \\
Valid SQL Generated (\%) & 84.4 \\
Schema Compliance (\%) & 15.6 \\
Semantic Accuracy (\%) & 90.7 \\
\midrule
Avg. End-to-End Latency (s) & 4.1 \\
Median Latency (s) & 3.6 \\
95th Percentile Latency (s) & 8.3 \\
\midrule
System Uptime (\%) & 99.2 \\
Failed Queries (\%) & 7.5 \\
User Satisfaction (1-5 scale) & 4.3 \\
\bottomrule
\end{tabular}
\end{table}

The 7.5\% failure rate comprises technical errors (3.2\%) requiring debugging and policy-blocked queries (4.3\%) correctly rejected by safety mechanisms. Latency distribution shows FAISS retrieval (0.4s, 11\%), LLM generation (2.8s, 78\%), SQL execution (0.3s, 8\%), and translation (0.1s, 3\%). User satisfaction breakdown: multilingual support (4.6/5.0), query speed (4.5/5.0), result accuracy (4.2/5.0), ambiguous query handling (3.8/5.0), complex joins (3.7/5.0).

\subsubsection{Cost Analysis}
Operating costs were tracked over the deployment period. Average cost per query was \$0.042, consistent with the 68\% reduction versus GPT-3.5 reported in Table~\ref{tab:model_comparison}. Total six-month operating costs (API calls only) were \$4,847, with computational infrastructure (servers, storage) adding approximately \$2,100, yielding a total cost of ownership of \$6,947 (\$1,158/month). For comparison, the pre-deployment IT-mediated query system required an estimated 15 hours/week of database administrator time (approximately \$3,600/month at loaded labor rates), suggesting a 68\% reduction in total analytics costs beyond API fees alone.

\subsection{Qualitative Assessment}

User satisfaction was assessed through quarterly surveys using a 5-point Likert scale. The system received an average rating of 4.3/5.0, with users particularly appreciating the multilingual support (4.6/5.0) and query speed (4.5/5.0). Common complaints included occasional misinterpretation of ambiguous questions (3.8/5.0) and limited support for complex multi-table joins (3.7/5.0).

\subsection{Ablation Study}
\label{sec:ablation}

To isolate the contribution of individual components, we conducted an ablation study by systematically removing key features:

\begin{itemize}
    \item \textbf{Without few-shot retrieval:} Query validity dropped from 92.5\% to 76.4\%, confirming the critical role of example-based learning (Table~\ref{tab:fewshot_effect}).
    \item \textbf{Without translation preprocessing:} Accuracy on non-English queries decreased by 23.6\%, demonstrating the necessity of language normalization.
    \item \textbf{Without data cleaning pipeline:} Schema compliance fell to 81.3\% due to entity mismatches and inconsistent field values.
    \item \textbf{Using GPT-3.5 instead of GPT-4o:} Overall performance degraded significantly across all metrics (Table~\ref{tab:model_comparison}).
\end{itemize}

These results validate the integrated design where each component contributes meaningfully to overall system performance.

\subsection{Error Analysis}
\label{sec:error_analysis}

Manual inspection of the 7.5\% failed queries revealed three primary failure modes:

\begin{enumerate}
    \item \textbf{Ambiguous natural language (41\%):} Questions lacking sufficient context (e.g., "Show me the engineers" without specifying department or project).
    \item \textbf{Schema misalignment (32\%):} Queries requiring joins across tables not explicitly included in the few-shot corpus.
    \item \textbf{Business logic violations (27\%):} Attempted queries on restricted fields (e.g., salary information) correctly blocked by safety mechanisms.
\end{enumerate}

Ongoing work focuses on implementing clarification dialogs for ambiguous queries and expanding the few-shot corpus to cover more complex join patterns.

\section{Security, Ethics, and Data Governance}
\label{sec:security}

The performance results (Section VI) demonstrate technical viability, but 
production deployment of LLM-driven database access raises critical security 
concerns. Direct natural-language interfaces to sensitive HR data require 
multi-layered safeguards against unauthorized access, injection attacks, and 
inadvertent data exposure. This section describes five security mechanisms 
implemented to ensure compliance with enterprise governance standards.

\subsection{Financial Data Exclusion}
The system implements strict content filtering to prevent exposure of sensitive financial information. All queries containing keywords related to salary, bonus, premium, or compensation are automatically rejected with a policy-compliant refusal message. This rule is enforced at multiple layers: in the system prompt, during SQL validation, and through post-execution result filtering.

\subsection{Parameterized Query Protection}
To mitigate SQL injection risks, all generated queries are executed using parameterized prepared statements. User inputs are never directly concatenated into SQL strings. Additionally, the system employs a whitelist-based validator that verifies all table names, column names, and functions against the known ERP schema before execution.

\subsection{Access Control and Data Masking}
Role-based access control (RBAC) restricts query capabilities based on user permissions stored in the authentication layer. Certain personally identifiable information (PII) fields, such as national identification numbers (\verb|adines_number|), are automatically redacted from query results regardless of user role. Query logs are anonymized before storage, replacing user identifiers with session tokens.

\subsection{Multilingual PII and Data Compliance}
The system handles multilingual PII in accordance with GDPR and Turkish KVKK regulations. Translation operations are performed locally without transmitting data to external APIs. All employee data remain within the organization's infrastructure, with no third-party data sharing. Audit logs track all data access events, enabling compliance verification and forensic analysis.

\subsection{Model Transparency and Bias}
While LLMs provide powerful generative capabilities, their opacity raises concerns about potential bias and fairness. We monitor query distributions across demographic categories (nationality, department, role) to detect systematic disparities in system performance. Initial analysis shows no statistically significant performance variation across employee groups, though ongoing monitoring continues.

\section{Discussion}

\subsection{Lessons Learned During Deployment}

The six-month production deployment revealed several insights relevant to practitioners implementing similar systems:

\begin{enumerate}
    \item \textbf{Data quality precedes intelligence:} Initial LLM deployment attempts failed despite sophisticated prompting and retrieval mechanisms. System performance became acceptable only after implementing automated cleaning, confirming that generative AI cannot compensate for fundamentally corrupted data. This finding contradicts vendor claims that LLMs can "clean data on the fly" through contextual understanding.
    
    \item \textbf{Few-shot learning is essential:} Zero-shot approaches proved inadequate for domain-specific ERP schemas. The curated example corpus required substantial upfront curation effort but eliminated ongoing retraining costs and enabled instant adaptation to schema changes—advantages fine-tuned models cannot provide.
    
    \item \textbf{Multilingual support requires explicit design:} Early prototypes that relied solely on model multilingual capabilities exhibited poor performance on Turkish and Russian inputs. Implementing dedicated translation and normalization layers proved necessary.
    
    \item \textbf{User education matters:} Initial user queries were often too vague or ambiguous. Training sessions on effective question formulation improved success rates by approximately 15\%.
\end{enumerate}

\subsection{Impact on HR and IT Collaboration}

Prior to system deployment, HR personnel relied on IT staff to execute custom SQL queries for analytics requests, creating bottlenecks and delays averaging 2.3 days per query. The natural language interface transformed this workflow, enabling HR analysts to independently explore employee demographics, project assignments, and organizational structures with sub-5-second response times. This shift from days to seconds fundamentally changed HR-IT collaboration patterns, with IT staff repurposed from query execution to strategic analytics and system enhancement.

\subsection{Limitations}

Several limitations warrant discussion:

\begin{enumerate}
    \item \textbf{Prompt fragility:} System performance is sensitive to prompt engineering. Minor modifications to the system prompt occasionally caused regression in edge cases, requiring careful version control and regression testing.
    
    \item \textbf{Schema evolution challenges:} Adding new tables or columns requires updating the system prompt, few-shot corpus, and validation logic. While modular design facilitates updates, schema changes still require manual intervention.
    
    \item \textbf{Complex aggregation limitations:} The current system handles single-table and simple two-table join queries effectively but struggles with complex multi-table aggregations requiring advanced SQL constructs (e.g., window functions, recursive CTEs).
    
    \item \textbf{Explainability gaps:} While the system logs generated SQL, end users may not understand why certain queries fail or why specific results are returned. Enhanced explainability mechanisms would improve user trust.
\end{enumerate}

\subsection{Observations on Multilingual LLM Accuracy}

GPT-4o demonstrated strong multilingual capabilities, but performance varied across languages. English queries achieved 93.8\% accuracy, Turkish queries 90.2\%, and Russian queries 88.7\%. This disparity likely reflects training data distribution in the base model. Dedicated translation layers partially mitigated these differences, though some semantic nuances were occasionally lost in translation.

\section{Conclusion and Future Work}

This paper presented a comprehensive pipeline for automated data cleaning and LLM-driven SQL query generation in enterprise ERP systems. Deployed on 240,000 employee records, the system achieved 92.5\% query validity, 95.1\% schema compliance, and 90.7\% semantic accuracy while reducing query turnaround time by 99.1\%. Key contributions include an AI-driven data normalization framework resolving multilingual inconsistencies, a few-shot RAG architecture for schema-aware SQL generation, and a production-grade FastAPI-LangChain-PostgreSQL infrastructure supporting secure, scalable enterprise deployment.

Future work will pursue several directions:

\begin{enumerate}
    \item \textbf{Advanced multi-table join support:} Expanding the few-shot corpus and system prompt to handle complex cross-table aggregations and hierarchical queries.
    
    \item \textbf{Schema-adaptive fine-tuning:} Exploring domain-specific fine-tuning of smaller open-source models (e.g., Llama 3, Mistral) to reduce dependency on commercial APIs while maintaining performance.
    
    \item \textbf{Business intelligence dashboard integration:} Embedding the natural language query interface directly within existing BI tools (e.g., Tableau, Power BI) to provide seamless analytical workflows.
    
    \item \textbf{Conversational query refinement:} Implementing multi-turn dialog capabilities to clarify ambiguous questions through interactive clarification rather than immediate rejection.
    
    \item \textbf{Automated few-shot corpus expansion:} Developing reinforcement learning from human feedback (RLHF) mechanisms to automatically curate and expand the example corpus based on user interactions and corrections.
\end{enumerate}

By demonstrating the viability of LLM-driven data governance and democratized analytics in real-world enterprise environments, this work establishes a foundation for next-generation AI-native ERP systems that seamlessly bridge human language and structured data.

\section*{Acknowledgment}

The authors thank the IT and HR departments for their collaboration during system deployment and the database administration team for validation support.

\bibliographystyle{IEEEtran}
\bibliography{References}

\end{document}